# Scene Synchronization for Real-Time Interaction in Distributed Mixed Reality and Virtual Reality Environments

Felix G. Hamza-Lup[1] and Jannick P. Rolland[1,2]

[1]School of Electrical Engineering and Computer Science
[2]School of Optics-CREOL
University of Central Florida
fhamza@cs.ucf.edu, jannick@odalab.ucf.edu

**Abstract**

Advances in computer networks and rendering systems facilitate the creation of distributed collaborative environments in which the distribution of information at remote locations allows efficient communication. One of the challenges in networked virtual environments is maintaining a consistent view of the shared state in the presence of inevitable network latency and jitter. A consistent view in a shared scene may significantly increase the sense of presence among participants and facilitate their interactivity. The dynamic shared state is directly affected by the frequency of actions applied on the objects in the scene. Mixed Reality (MR) and Virtual Reality (VR) environments contain several types of action producers including human users, a wide range of electronic motion sensors, and haptic devices. In this paper, the authors propose a novel criterion for categorization of distributed MR/VR systems and present an adaptive synchronization algorithm for distributed MR/VR collaborative environments. In spite of significant network latency, results show that for low levels of update frequencies the dynamic shared state can be maintained consistent at multiple remotely located sites.

# 1. Introduction

Technological advances in optical projection and computer graphics allow users in virtual environments to span the Virtuality Continuum (Milgram & Kishino, 1994) from real worlds to entirely computer generated environments with the opportunity to also augment their reality with computer generated three-dimensional objects (Billinghurst, Kato, Kiyokawa, Belcher, & Popyrev 2002; Davis, Rolland, Hamza-Lup, Ha, Norfleet, Pettitt, & Imielinska, 2003). Particularly, the distribution of three-dimensional objects at remote locations allows efficient communication among remotely located participants. For an effective collaboration, all users of the system must be able to see the effects of the interaction at the same time. Moreover, the consistency of the dynamic shared state is directly affected by the frequency of actions applied on the virtual objects in the scene.

As an example, consider the following application scenario. A surgeon is in his office analyzing a 3D model of the mandible of one of his patients. He would like to discuss the surgical procedure that will follow shortly with one of his colleagues, whose office is in another building. As part of the discussion, they have to analyze the 3D model of the patient's mandible. They use the 3D distributed visualization platform implemented on the hospital's local area network. For stereoscopic visualization, each office is equipped with a head mounted display and a sensing glove (e.g. P5 Glove from Essential Realty Inc.). In this scenario, the distributed visualization platform allows one user to modify the position and orientation of the 3D model from a mouse-driven graphical user interface (GUI) or through the sensing glove.

There are two problems that arise from this scenario. The first problem is related to the network latency. As one of the users manipulates the 3D model, the network latency desynchronizes their common viewpoints. Moreover, since network jitter is also present, the position/orientation drift among the views increases in time while the users are not aware of the inconsistency of their viewpoints.

The second problem pertains to the nature of the interaction with the objects in the shared scene. The 3D model can be either manipulated from a graphical user interface through discrete and predictable actions, or using the glove-like peripheral device through continuous and relatively unpredictable actions. The human user acting on the graphical user interface through the mouse, for example, cannot exceed a certain frequency of actions mainly because of his motor reaction time. At the same time, since the position and the orientation of the object are set through the interface, predictable actions are applied to the object (e.g. by pressing the GUI's "Rotate around OX axis" button). In contrast to the GUI, the glove-like peripheral device is usually tracked at high frequencies (e.g. a P5 glove has an optical tracking system attached that has a refresh rate of 60 Hz) and is going to capture the user actions at a higher frequency (e.g. even the user's insignificant unintentional hand shaking will be captured). As a result, we have two types of interaction with the 3D model that have distinct patterns.

While the network latency problem is well known in distributed collaborative VR, the second problem is more subtle and requires further analysis. Based on the above observations, we propose a novel criterion for categorization of distributed MR/VR applications as well as an adaptive synchronization algorithm that takes into account the network latency magnitude. Other factors that affect the synchronicity of a distributed MR/VR system, besides network latency, are differences in the hardware architectures over the system's nodes, hardware buffering, and software system delays (Swindells, Dill, & Booth, 2000). In this paper, however, we assume that the most relevant factor is the network latency. The proposed algorithm ensures superior synchronization of the shared scene for remotely located participants by compensating the network latency.

The rest of the paper is organized as follows. Section 2 discusses related work while section 3 describes the criterion for categorization of distributed MR/VR applications, the adaptive synchronization algorithm employed to compensate for the network latency and an adaptive strategy for measuring network delays. Section 4 presents the integration of the algorithm within a framework and a method for synchronization assessment. Section 5 focuses on the experimental results, confirming our theoretical analysis. Finally, section 6 concludes the paper and identifies areas of future improvement and research.

## 2. Related Work

Previous work points to the challenges of managing the dynamic shared state and the consistency-throughput tradeoff (Singhal & Zyda, 1999). Research efforts have concentrated on the development of middleware on top of client-server models for distributed data sharing through remote method calls using an object oriented approach (MacIntyre & Feiner, 1998) (Tramberend,1999). A drawback of building the middleware frameworks on the distributed object model is the additional latency caused by the software layers, leading to non-real time behavior and poor scalability.

Recently, Schmalstieg and Hesina presented an augmented reality (AR) framework, Studierstube, which uses a distributed shared scene graph to keep track of the actions applied on the shared scene by multiple users (Schmalstieg & Hesina, 2002). The authors show that multiple concurrent operations on the objects in the scene may lead to highly inconsistent views. As communication delay increases, the inconsistency among remote users of the system increases. Therefore, synchronization strategies must be employed for maintaining a consistent view among the users of the system.

In the DEVA3 VR system (Pettifer, Cook, Marsh, & West, 2000), each entity is composed of a single "object", which represents what the entity does, and a set of "subjects" which represents how the entity looks, sounds and feels. Partial/weak consistency is maintained by updating an object position only when the subject has moved a certain distance from its previously synchronized location. Another synchronization approach is available in MASSIVE-3 (Greenhalgh, Purbrick, & Snowdon, 2000), a predecessor of the HIVE VR system. The updates (Sung, Yang, & Wohn, 1999) in MASSIVE-3 are effective combinations of centralized updates and ownership transfers. The centralized update approach reduces the scalability of the system.

Finally, a common approach to synchronization is to timestamp the outgoing messages using an external clock synchronization method, such as Network Time Protocol (NTP). Such approach provides accuracies typically within milliseconds on local area networks and up to tens of milliseconds on wide area networks relative to Coordinated Universal Time using a global positioning service receiver, for example. However, to investigate the latency between two nodes in the network, the difference between the time the packet was stamped at the source and at the destination must be computed. Assuming a similar path from source to destination and back, we shall show how a simple ping-like probe gives remarkable results.

None of the reviewed approaches exploit the network latency magnitude to maintain a consistent shared scene. The novelty of the synchronization algorithm presented in this paper consists of the compensation for both the network latency and the network jitter. Moreover we propose a novel criterion for characterizing MR/VR distributed applications based on the patterns of the actions applied on the virtual components of a scene.

## 3. Classifying Distributed MR and VR Applications

It is important, as a first step in a distributed MR /VR application, to ensure that each participant has the appropriate resources for the shared experience. If the distributed application must ensure interactive speed, the appropriate resources must be available at specific time instances. MR/VR applications require a set of virtual 3D objects. These objects usually have a polygonal representation which facilitates fast rendering. However, this representation requires a large storage space. Distributing these 3D objects on a local area network while the distributed MR/VR application is running, negatively affects the interactivity of the application. A solution to this problem is to have all the MR/VR scene components downloaded asynchronously at each node before the interactive application starts. Thus, the only data that circulates among the nodes is the data generated by the users that interact with the virtual objects in the shared scene. We now define a novel criterion for classifying applications based on the action frequency patterns.

### 3.1 Action Frequency Patterns: High vs. Low Frequency

There are several ways of interacting with virtual objects. In distributed MR/VR applications, the position and orientation of the virtual objects can be dictated either by the users through graphical user interfaces or by sensors attached at different locations in the real environment. Moreover, the data produced through these devices has to be distributed such that each participant can update its local scene to have a consistent view of the shared scene. The distribution is usually done through a delay prone environment (e.g. a network), which has a lower bound on the latency. Intuitively, two cases can be established; the actions frequency is either lower or higher than the network delay between two interacting nodes. To study these issues, we define two frequencies in the following paragraph: the upshot frequency and the action frequency. We than show how these frequencies can be used to categorize distributed interactive MR/VR applications as *high* or *low* frequency applications. Without loss of generality and to avoid confusion between a node and a participant, throughout the paper, we assume that each physical node in the distributed system on which the MR/VR application runs has one and only one human participant associated. For the remainder of the document, the term "participant" and "node" will be used interchangeably.

Let $\delta$ be the average network delay between two participating nodes. We assume, for now, that there is no jitter. The jitter compensation is described in section 3.3. Let us consider a distributed MR/VR application that allows interaction with $m$ virtual 3D objects. The shared scene produced by this application must be displayed at all the participating nodes. Let $n$ be the number of participating nodes. We assume that all virtual objects are rigid and we restrict the actions applied on them to rotations and translations. The discussion can be further extended to arbitrary affine transforms and non-rigid, deformable 2D/3D objects.

During the application execution, the participants interact with the virtual objects in the scene. Each participant interaction can be seen as a sequence of actions applied on the objects in the scene. An action is identified by a name, a direction described by a vector, and a velocity. Each time one of these attributes changes a new action is born. We assume that the acceleration is zero but the model can be easily extended to higher-order derivatives with additional computational expense. Since real-life interactions are spontaneous, the action duration is not known when the action is applied. The action duration is known only after the next action is initiated (i.e. when one of the action attributes changes), and can be computed as a difference between the current action initiation time and the previous action initiation time.

We define the *action frequency* (denoted by $v_k$) of a node $k$, as the number of actions performed by node $k$ on one object in the shared scene per unit time. The action frequency is measured in actions per second and can be estimated in the following way. Let $b_{jk}$ be the number of actions applied on the object $j$ in the scene by participant $k$ during $\Delta t$. Participant $k$ may interact with any object in the shared scene. The total number of actions applied by $k$ on all $m$ objects in the scene during $\Delta t$ will be $\sum_{j=1}^{m} b_{jk}$. An estimate of the action frequency for participant $k$ (i.e. $v_k$) can be obtained by computing the average number of actions applied by $k$ on an object in the scene, which is given by:

$$v_k = \frac{\sum_{j=1}^{m} b_{jk}}{m \cdot \Delta t} \qquad (1)$$

The estimation can be done for each participant in the distributed application.

Furthermore, let $v_0$ be the *upshot frequency* between two participants defined as $v_0 = \dfrac{1 \text{ action}}{\delta}$, with actions per second as measurement units. The upshot frequency is dependent on the network delay between two participating nodes. For example, if $\delta$ is 100 milliseconds, the upshot frequency between the nodes will be 1/0.1 or 10 actions per second. Computing the average network delay between each pair of participants, the corresponding upshot frequency ($v_0$) can be computed.

Based on these definitions two cases can be distinguished: the action frequency is less than the upshot frequency, *($v_k < v_0$)*, and the action frequency is greater than or equal to the upshot frequency, *($v_k \geq v_0$)*.

To illustrate the discussion above, consider a simple case in which there is only one virtual 3D object in the shared scene and two participants, node *X* and node *Y*, located on a local area network. Node *X* can change the object orientation by applying arbitrary rotations around the object coordinate axes. Since it is a distributed application, the shared scene must be maintained consistent (i.e. both participant should see the same orientation for the object).

The first case *($v_k < v_0$)* usually corresponds to a distributed MR/VR application involving either low update frequency devices (as compared with the upshot frequency) or users who perform actions on the objects in the shared scene through a graphical user interface. The fastest human-computer response time includes perceptual, conceptual, and motor cycle time, which adds up to an average of about 240ms (Eberts & Eberts, 1989). Under the later assumptions, distributed applications that fall in this category should not be deployed on a network that has an *upshot frequency* lower than 4.16(actions/second), in other words the network delay has to be lower than 240ms. In most of the cases when the distributed MR/VR application is deployed on a local area network, the frequency of the actions applied by a human participant through a GUI on the objects in the shared scene will be below the upshot frequency. Moreover, some actions will generate continuous movements that can be predicted. For example, the user might spin an object for an indefinite time period with a specific velocity around a specific axis. In this case, $v_0/v_k$ tends to infinity and once the nodes are synchronized no additional network traffic is necessary until another action is applied. The drift can be accurately computed if we know the network delay and the action (e.g. rotation) velocity.

We emphasize this scenario with an example. A timing diagram is shown in Fig.1 that contains two actions applied by node X on an object in the shared scene and the propagation of these actions to another participant, node Y. The first action, $a_1$, takes 14 time units, the second action, $a_2$, takes 8 time units, and the network delay between nodes *X* and *Y* is 1 time unit. The synchronization algorithm proposed in section 3.2 accounts for the network latency. In Fig.1 the shaded areas represent the time intervals when *X* and *Y* are synchronized.

**<<Fig.1 Action frequency is less then the Upshot frequency ($v< v_0$), $\delta =1$.>>**

The second case *($v_k \geq v_0$)* corresponds to MR/VR applications containing a fast updating device like a tracking system or a high latency network connection. In this case, a sequence of actions might take place at node *X* before node *Y* is notified about the first action in this sequence. This scenario can be described with a simple example. Fig. 2 is a timing diagram that contains 10 actions and their respective durations: actions $a_1, a_2, a_3, a_5, a_6, a_7,$ take 1 time unit; action $a_4$ takes 2 time units, and actions $a_8, a_9, a_{10}$ take 3 time units. The network delay between nodes $X$ and $Y$ is 3 time units.

<<**Fig.2 Action frequency is greater or equal with the Upshot frequency ($v \geq v_0$), $\delta = 3$.**>>

In this case, high quality of synchronization cannot be reached since node $Y$ will continuously try to "*catch up*" with node $X$. By the time node $Y$ has compensated for the drift, node $X$ has applied new actions on the object. Node $Y$ is not aware of those actions at that time. Position prediction techniques, like the Position History-Based Dead Reckoning protocol, (Singhal & Cheriton, 1995) can be employed as approximate solutions in this case.

### 3.2 Adaptive Synchronization Algorithm

The adaptive synchronization algorithm proposed is targeted towards MR/VR applications which fall in the first category *($v < v_0$)* as defined in the previous section. Applications that meet this criterion will become increasingly available as low latency networks and optical routing become widespread. Moreover, the algorithm assumes a homogenous distributed system. The experiments described in section 5 show that the implementation gives satisfactory results with nodes having slightly different rendering capabilities and similar network connections. A practical example of a MR/VR distributed applications were the algorithm was used is described in section 4.

To control the position and orientation of the objects in the shared scene, each 3D object has a control packet object (CPO) associated with it. The CPO is a software entity, an instance of a class, and contains information about the position and orientation of the 3D object in the scene as well as information regarding the actions associated with it. The small size of the CPO (i.e. several KB) ensures a very low propagation delay, which allows the development of scalable, distributed interactive applications on a local area network. As the CPOs are transmitted through the network, the adaptive synchronization algorithm uses their information to synchronize the shared scene among different participants. The information carried by the CPOs is distributed to each participating node, allowing them to compensate for the network latency.

Each object in the shared scene has an associated lock. To change the position and orientation of a set of objects, a node must acquire their locks first. Furthermore, for that set of objects, the node that acquires the "locks" will act as a server, while the other nodes act as clients receiving new position information for the objects in the shared scene. In the following discussion we assume that one node has acquired the locks of all the objects in the scene and acts like a server while the others receive updates. More complex scenarios (e.g. several server nodes handling disjunctive subsets of objects in the shared scene) can be decomposed in simpler scenarios like the one above.

We define the position/orientation ***drift value*** for a particular object $j$ and a particular node $k$ as the product between the action velocity (*meters/second or degrees/second*) applied on the object and the network delay from the server to client node $k$. If we correct the position/orientation of the object displayed on node $k$ according to the drift value, we can achieve ideal synchronization between the server and node $k$. The drift computation and the associated correction applied independently on each node is the essence of the adaptive synchronization algorithm.

Let $m_\tau$ be the number of virtual objects in the shared scene and $n_\tau$ the number of participating nodes at a given time $\tau$. A drift matrix $D(m_\tau, n_\tau)$ associated with the distributed system at time $\tau$ may be defined as:

$$D(m_\tau, n_\tau) = S \cdot T^t \qquad (2)$$

where $S$ and $T$ are both column vectors, one containing the action velocities for each object currently in the shared scene, and the other the network delays from each participating node to the current server. $T^t$ represents the transpose of $T$. The action velocity is extracted from each object's CPO, while the network delay is measured by each node using an adaptive probe that computes the round trip time from the node to the server as discussed in section 3.3. The action velocities vector, $S$, is stored locally at each node and updated when the scene changes.

A decentralized computational approach strips the drift matrix in $n$ column vectors, called *drift vectors*, which contain the drift values of all the objects in the scene for a particular node. The drift vectors are updated when a new 3D object is inserted or removed from the shared scene by adding or removing respectively the entry associated with the new object from all nodes. The drift vectors are also updated when the users perform actions on the objects in the shared scene. Whenever an action is applied on an object (e.g. a rotation), the CPO associated with that object is broadcasted to all the nodes. The information from the CPOs is the first component used for synchronization. The second component accounts for network delays. At regular intervals each node "pings" the server to estimate an average network delay and computes the drift vectors associated with the objects in the scene as a product between the average network delay and the action velocity of each object. Each delay measurement between a node and the server triggers the node's drift vector update.

The main part of the Adaptive Synchronization Algorithm is further described in pseudocode. The *ComputeNodeDelay()* function returns the delay associated with the connection between a client node and the server. The *UpdateActions()* function updates the actions' attributes (e.g. name, velocity) associated with the objects in the scene. The *UpdateDrift()* function updates each node with its drift values for the objects in the scene. Three boolean variables are used: *changedScene* that accounts for the changes in the scene, *newClientRequest* which is set when a new client has joined, and *trigger* which tracks the network behavior as described in section 3.3. Finally the function *BroadcastChanges()* ensure correct scene updates among the nodes of the system and the server. Each node's scene is synchronized with the server. Hence, each node is synchronized with all the other nodes.

> *Algorithm: Adaptive Synchronization*
> *Output: Synchronized shared scenes for a distributed interactive VR/MR application.*
> *Client side:*
> > *Initialization:*
> > > $T_n \leftarrow$ *ComputeNodeDelay()*
> > > $S_n \leftarrow$ *UpdateActions();*
> > > $D_n \leftarrow$ *UpdateDrift()*
> > > *UpdateLocalScene();*
> > *Main:*
> > > *if (trigger)*
> > > > $T_n \leftarrow$ *ComputeNodeDelay()*
> > > > $D_n \leftarrow$ *UpdateDrift()*
> > > *end if*
> > > *if (changedScene)*
> > > > $S_n \leftarrow$ *UpdateActions()*
> > > > $D_n \leftarrow$ *UpdateDrift()*

```
                                end if
            Server side:
                for ever listen
                    if (newClientRequest)
                        SendToClient(S_n);
                    end if
                    if (changedScene)
                        BroadcastChanges();
                    end if
                end for
```

### 3.3 Fixed Threshold vs. Adaptive Threshold Synchronization

As the traffic in the network changes, the round trip times between different nodes vary. To achieve the best synchronization possible among collaborating nodes, delay measurements must be triggered at different rates. The measurement rates must follow the network jitter behavior. The goal is to obtain an accurate estimate of the average delay from each client node to the server node. An average round trip time (RTT) can be obtained by sending "ping" messages to the newly arrived node when it joins a group. Half of this delay represents an average delay from the node to the server.

The synchronization algorithm uses two approaches to *trigger* the information collection. In the first approach, at regular time intervals (e.g. every second), using the Internet Control Message Protocol (ICMP), a node opens a raw socket and measures the RTT to the current server. We denoted this approach as the "fixed threshold" approach.

Gathering the RTT data imposes additional overhead at the server and additional network traffic. Moreover these measurements are not required if the network jitter is very low. An alternative approach consists of *adaptively* triggering the delay measurements for each node, based on the delay history, which better characterizes the network traffic and the MR/VR distributed application. In the adaptive approach, a fixed threshold is initially used at each node to build the delay history, denoted $H_p$. The delay history is a sequence of $p$ delay measurements $h_i$ where $i=1,p$ (e.g. in the implementation we have chosen $p$ to be 100). Furthermore, let $h_{mean}$ and $\sigma$ be the mean and standard deviation of $H_p$, respectively.

Let $h_0$ be the most recent delay, i.e. the last number in the $H_p$ sequence, and $\gamma_0$ the current frequency of delay measurements, expressed as the number of measurements per second. The adaptive strategy is to decrease $\gamma_0$ if $h_0 \in [\ h_{mean} - \sigma,\ h_{mean} + \sigma\ ]$ and to increase $\gamma_0$ if $h_0$ does not belong to this interval. The adaptive synchronization algorithm has been embedded in DARE, a Distributed Augmented Reality Environment (Hamza-Lup, Davis, Rolland, & Hughes, 2003).

**4. Distributed Artificial Reality Environment**

DARE is a framework which uses AR and VR techniques to improve human-to-human interaction by enhancing the real scene that a person sees with 3D computer generated objects. Applications built on this platform range from distributed scientific visualization to interactive distributed simulations and span the entire Virtuality Continuum (Milgram & Kishino, 1994).

3D Remote Collaborative Scientific Visualization is an application built on this framework. In this application, the first step consists of determining the resources (i.e. the 3D models) needed for visualization. The 3D models are downloaded asynchronously to each of the participating sites (nodes), allowing the distributed visualization to have an interactive behavior. The data that is synchronously sent is the data associated with the users' interactions on the shared scene. This data is embedded in CPOs, which have very small propagation time.

**4.1 Hardware System Components**

Each DARE node consists of a head-mounted projection display (HMPD) (Hua, Ha, & Rolland, 2003), a Linux based PC, a quasi-cylindrical room, called an Artificial Reality Center (ARC), and walls covered with retroreflective material (Davis, Rolland, Hamza-Lup Ha, Norfleet, Pettitt, & Imielinska., 2003). The algorithm performance is slightly affected by the platform where the implementation is deployed. The Windows™ operating system is challenging to control at a fine level therefore we have deployed the algorithm implementation on a Linux based platform.

As users wearing HMPDs enter the ARC, they gradually start immersing themselves in virtuality. Initially, the users' reality is augmented with 3D computer-generated objects. These virtual objects may appear to multiple users if they share the same scene. Users can also interact with the 3D objects. Using a graphical user interface with 3D pointing capabilities, they can manipulate these objects and they can point in the virtual space to different parts of the objects (Fig.3).

<<Fig.3  a) 3D pointing     b) Local collaboration     c) Remote user>>

Several ARC rooms can be interconnected on a local area network allowing remote stereoscopic visualization as described in Fig.4. These Networked Open Environments allow remote collaboration through distributed applications that span the entire virtuality continuum.

<<Fig.4 NOEs ARCs (Networked Open Environments with Artificial Reality Centers)>>

**4.2 A Method for Synchronization Assessment**

To assess the efficiency of the synchronization algorithm, the amount of orientation /position drift between the pose of a 3D object on two nodes must be measured. In this section the discussion is focused on the assessment of the orientation drift. A similar assessment can be done for the object's position. We use two nodes (participants) sharing the same virtual 3D scene, with one acting as a server and the other as a client. A graphical user interface is available at the server site, which allows the user to change the object *orientation* by applying rotations around the Cartesian axes. The participant generates events from the interface, and each time an event is generated, the object's orientation at both sites is recorded. Because of the network latency, different vectors at each node will describe the orientation of the object. The rotations can be easily expressed using quaternion notation.

Let $q_s$ express the rotation of an object at the server node and let $q_c$ express the rotation of the same object at the client node. Both nodes render the same virtual object and the object should have exactly the same orientation. To quantify the difference between the orientations of the object on two different nodes, we can compute the correction quaternion $q_E$ between the nodes every time the user triggers a new action. The correction can be expressed as

$$q_s = q_E q_c \quad (3)$$

and thus

$$q_E = q_s q_c^{-1}. \quad (4)$$

The quaternion $q_E$ may be expressed as

$$q_E = (\omega_E, \vec{v}_E) = (\cos(\frac{\alpha}{2}), \sin(\frac{\alpha}{2})(x\hat{i}, y\hat{j}, z\hat{k})), \quad (5)$$

where

$$\alpha = 2\cos^{-1}(\omega_E). \quad (6)$$

The angle α represents the drift between the orientations of a 3D object displayed by both nodes.

## 5. Experimental Setup and Results

To evaluate the performance of the algorithm, we first calculated the network latency using a latency measurement probe on a 100 Mbps LAN. The average round trip time for this setting was 1.5 ms. To investigate the effects of the network latency, we repeated the experiments at different action velocities, given that the drift value for an object is the product between the action velocity and the network latency (as defined in section 3.2). To prove the scalability of the system regarding the number of participants, two sets of experiments were performed. The first set contained two nodes: one acting as a client and the other one as a server. The second set contained 5 nodes, one acting as a server and the other 4 as clients.

### 5.1 Two Nodes Setup: Network Latency Analysis

MIT Press

Running the distributed visualization with and without the synchronization algorithm, we can assess the effectiveness of the algorithm. Fig.5 provides a plot of the orientation drift angle ($\alpha$) for various action velocities before synchronization. The actions in this case were random rotations of a virtual object around its coordinate axis with the angular velocity of 10, 50 and 100 degrees per second.

<<Fig. 5 The angular drift ($\alpha$) without synchronization for different angular velocities>>

The plot shows that as the action velocities increased, the drift also increased, as expected, and the magnitude of the drift reached over 140 degrees after 24 actions, for high action velocities. Overall, the drift increased in time as more and more actions were applied on the object in the shared scene. The sudden drops in the drift were caused by the compensating factor of the random rotations (e.g. clockwise followed by counterclockwise rotations of the object around the same axis). The drifts compensated each other to some extent.

<<Fig. 6 The angular drift ($\alpha$) with synchronization for different angular velocities>>

The synchronization module activation caused a significant decrease in the drift as shown in Fig.6, where the maximum drift value is two orders of magnitude lower (i.e.3.1 degrees). To maintain the readability of the plots, we have changed the scale on the vertical axis of Fig.6 because the drift values were much smaller compared with the ones in Fig.5.

As the action velocity increased, the drift oscillation amplitude also increased. However in the worst case (defined by action velocities of 100 degrees per second) the drift value was maintained at an average of 2.4 degrees. Furthermore, the average drift value had almost a constant value during the simulation. Increasing the action velocity to 100 degrees per second, on a network having 1.5 ms latency, would be equivalent to running the distributed application on a network having 15 ms latency using action velocities of 10 degrees per second. Thus, the algorithm may be applicable to distributed MR/VR applications running on a wide area network where participants are separated by network latencies higher than 15 ms.

**5.2 Five Nodes Setup: Scalability Analysis**

To test the scalability of the algorithm, a five node setup was implemented. This setup allowed five remote participants to be part of the distributed interactive application. One of the nodes ran the server process and the participant at this node was able to change the position and orientation of the virtual objects in the scene. In the current implementation, the other 4 participants did not interact with the scene. They were only able to visualize the virtual scene. We recorded the orientation of one virtual object while the participant on the server node applied rotations on the object with different speeds. The other four nodes ran client processes and they were able to visualize the same virtual scene. Every new event generated from the server node triggered an orientation update on the virtual object on all the other nodes. At the same time, the current orientation was recorded in memory on each node. The results were written to a file only once at the end of the simulation to reduce the intrusiveness. We tested a fairly heterogeneous configuration of nodes. The network cards on all nodes allowed 100Mbps connections. The table below contains a brief specification of each node's hardware components.

Table. 1 Hardware systems attributes

| Node no. | Arch. | CPU (GHz) | RAM (MB) | Video card (GeForce) |
|---|---|---|---|---|
| 1 | Desktop | 1.5 AMD | 1024 | 4 Ti4600 |
| 2 | Desktop | 1 P3 | 1024 | 2 Mx |
| 3 | Desktop | 1.7 P4 | 512 | 4 Mx 440 |
| 4 | Desktop | 1.7 AMD | 1024 | 4 Ti4600 |
| 5 | Laptop | 2 P4 | 1024 | 4 Go440 |

In the first stage, the simulation was executed without synchronization and at different action velocities. Fig.7 presents a plot of the angular drifts for different action velocities for each client node. The legend for Node 2 applies to Node 3, 4 and 5. Node 1 was acting as a server and was used as a reference for the drift computation. As in the first set of experiments, the results show that the drift increased as the action velocity increased. The drift variation over different nodes was caused by the hardware heterogeneity of the nodes.

**<<Fig. 7 The angular drift ($a$) without synchronization for different angular velocities on different nodes>>**

The second stage of the simulation was executed with the synchronization module active and at different action velocities. As the action velocity increased, it negatively affected the drift correction. However in all cases, the average drift angle at 100 degree/second action velocity did not exceed 3.5 degrees. Furthermore over all the nodes, the drift average was 2.9 degrees. Fig.8 illustrates the drift variations over different nodes with the synchronization module active. The legend for Node 2 applies to Node 3, 4 and 5. Node 1 was acting as a server and was used as a reference for the drift computation.

MIT Press

<<Fig. 8 The angular drift (*α*) with synchronization for different angular velocities on different nodes>>

The current client-server architecture on which the algorithm was deployed seems to introduce the disadvantage of data centralization. However, in our approach the majority of the computation is distributed among participating nodes. Each node renders its own scene and computes its own drift value. The only burden on the server node, which increases with the number of nodes, is the reply to each delay measurement message sent by a client node. The adaptive approach for triggering the network delay measurements described in section 3.3 has a positive impact on the scalability of the applications deployed on a stable network infrastructure. On the other hand, if the latency of the network infrastructure varies, the frequency of measurements triggered by each client node increases. If the number of participants also increases, the server might become ping flooded. Strategies in the category of ping flood protection might be employed in this case, which will limit the number of participants to the MR/VR collaborative application.

We shall now define a metric analyzing the relationship between the number of nodes in the system and the drift values. Assuming that the algorithm is activated let $\psi_i$ be the average drift value over all the nodes, when i+1 nodes are in the system. Without loss of generality, let us consider an action velocity of 100 degrees per second. In the case of a two node setup, results show that the average drift is $\psi_1$ equal 2.4 degrees, while in the case of a five node setup, the average drift is $\psi_4$ equal 2.9 degrees. An algorithm with low degree of scalability would have at least a linear increase in drift, i.e. $\psi_n$ equal n* $\psi_1$. On the other end, a high degree of scalability would mean $\psi_n \approx \psi_1$.

Using this metric, in the five node setup, a low degree of scalability would translate to $\psi_4$ equal $4*\psi_1$ or 9.6 degrees. However, the experimental results show that $\psi_4 \approx \psi_1$. Thus, the algorithm gives promising results in terms of scalability regarding the number of participants.

## 6. Conclusions and Future Work

An effective, distributed MR/VR collaborative environment that supports remote, real time interactions would allow users to approach their medical, engineering, or scientific data as a team, but with each participant holding a unique perspective. This may lead to startling observations and enhanced creativity from the participants, as a result of the synergy that develops among a group of people working together as if they are physically present in a common space.

In this paper we have introduced a novel criterion for categorization of distributed MR/VR applications based on the action frequency patterns. Such a criterion will help distributed collaborative environments designers to better match the application attributes with the network parameters.

In support of our criterion, we have presented and analyzed an adaptive synchronization algorithm which addresses the network latency problems in distributed MR/VR applications. The fundamental property of our design is that the algorithm takes into account and compensates for the network latency. The decentralized computation approach (i.e. independently at each node) for the drift values improves the system scalability and its real-time behavior. The algorithm is highly efficient when the network upshot frequency is higher than the user's action frequency. We believe in the widespread of such distributed applications as low latency optical networks and optical routing become increasingly available.

The next phase of our research will involve testing the current algorithm over different types of networks. We also plan to investigate the case when high frequency update sensors are connected in the distributed system (e.g. tracking systems and haptic devices). We are investigating the possibility of eliminating disconcerting jumps in the object position and orientation using interpolation and dead reckoning algorithms on a wide area network.

**7. Acknowledgements**

We thank our sponsors the NSF/ITR: IIS-00-820-16, the Link Foundation, and the US Army Simulation, Training, and Instrumentation Command (STRICOM) for their invaluable support for this research. We also thank Eric Clarkson, Larry Davis and Georgiana Hamza-Lup for stimulating discussions about quaternions. Finally, we thank Charles Hughes, Kien Hua, and Blair MacIntyre for motivating discussions about this area of research.

MIT Press